# Title: Minimally Interactive Segmentation of Soft-Tissue Tumors on CT and MRI using Deep Learning

Author list: Douwe J. Spaanderman, MSc[1]; Martijn P. A. Starmans, PhD[1]; Gonnie C. M. van Erp, MSc[1]; David F. Hanff, MD, PhD[1]; Judith H. Sluijter, MSc [1]; Anne-Rose W. Schut, MD[2,3]; Geert J. L. H. van Leenders, MD, PhD[4]; Cornelis Verhoef, MD, PhD[2]; Dirk J. Grünhagen, MD, PhD[2]; Wiro J. Niessen, PhD[5]; Jacob J. Visser, MD, PhD[1]; Stefan Klein, PhD[1]

[1]Department of Radiology and Nuclear Medicine, Erasmus MC, Rotterdam, the Netherlands
[2]Department of Surgical Oncology, Erasmus MC Cancer Institute, Rotterdam, the Netherlands
[3]Department of Medical Oncology, Erasmus MC Cancer Institute, Rotterdam, the Netherlands
[4]Department of Pathology, Erasmus MC Cancer Institute, Rotterdam, the Netherlands
[5]Faculty of Medical Sciences, University of Groningen, Groningen, The Netherlands

Institute address: Erasmus MC, Dr. Molewaterplein 50, 3015 GD Rotterdam, the Netherlands.

**Abbreviations**

CNN  convolutional neural network
DSC  Dice similarity coefficient
EGD  exponentialized geodesic distance
FS   fat saturated
RECIST response evaluation criteria in solid tumors
ROI  region of interest
STT  soft-tissue tumor
TCIA The Cancer Imaging Archive

## Abstract

Segmentations are crucial in medical imaging to obtain morphological, volumetric, and radiomics biomarkers. Manual segmentation is accurate but not feasible in the radiologist's clinical workflow, while automatic segmentation generally obtains sub-par performance. We therefore developed a minimally interactive deep learning-based segmentation method for soft-tissue tumors (STTs) on CT and MRI. The method requires the user to click six points near the tumor's extreme boundaries. These six points are transformed into a distance map and serve, with the image, as input for a Convolutional Neural Network. For training and validation, a multicenter dataset containing 514 patients and nine STT types in seven anatomical locations was used, resulting in a Dice Similarity Coefficient (DSC) of 0.85±0.11 (mean ± standard deviation (SD)) for CT and 0.84±0.12 for T1-weighted MRI, when compared to manual segmentations made by expert radiologists. Next, the method was externally validated on a dataset including five unseen STT phenotypes in extremities, achieving 0.81±0.08 for CT, 0.84±0.09 for T1-weighted MRI, and 0.88±0.08 for previously unseen T2-weighted fat-saturated (FS) MRI. In conclusion, our minimally interactive segmentation method effectively segments different types of STTs on CT and MRI, with robust generalization to previously unseen phenotypes and imaging modalities.

# Introduction

Soft-tissue tumors are rare tumors with a broad range of differentiation that can occur in a large variety of locations in the body. STT progression is highly variable across patients (1,2). The 3D delineation of STTs, i.e. segmentation, is needed for various purposes, such as targeted (neo)adjuvant radiotherapy planning (3), computation of quantitative imaging biomarkers (radiomics) (4-7), and calculation of Response Evaluation Criteria In Solid Tumors (RECIST) (8). Currently, these segmentations would have to be made manually which is a substantial burden on the physician's time, drives healthcare costs, and is observer dependent. Therefore, there is a need for more time-efficient, automated segmentation methods in clinical practice.

Fully automatic segmentation methods using deep learning have shown to be successful in various applications in medical imaging (9). However, their adoption in STTs has been limited due to the vast range of STT phenotypes, locations, and imaging modalities (10), which makes it difficult to train a fully automatic segmentation method that generalizes across all STT patients (10,11). A potential solution could be to allow a minimal amount of manual interaction, leveraging the radiologist's knowledge to guide the segmentation and thereby improve generalizability, while maintaining practical efficiency in a clinical setting (12).

The aim of this work was to develop and evaluate a minimally interactive deep-learning method for STT segmentation on CT and MRI. To this end, we adopted a previously proposed framework (12), optimized the methodology for STT segmentation, and trained it on a heterogeneous public dataset. For comparison, we also implemented a fully automatic segmentation method using the state-of-the-art nnU-Net framework (9). Both methods were compared to manual reference segmentations on independent test data. Finally, we conducted volume and diameter measurements to analyze the use of minimally interactive segmentation for clinical measurements.

## Materials and Methods

### Study Sample

In this study, we used the only two publicly available datasets including STTs. The characteristics of the datasets are described in **Table 1, Table S1** and **Supplementary Materials**.

For model development and validation, we used all STTs from the public, retrospective, multi-center WORC database, which includes 514 patients with either CT or T1-weighted MRI **(Figure S1)** (13). The WORC database includes reference tumor delineations obtained through manual segmentation on the T1-weighted MRI or CT scans by various clinicians under supervision of musculoskeletal radiologists (4-5 years of experience). For both CT and MRI, we split the dataset stratified on STT phenotype, using 80 percent for model development and 20 percent for validation, denoted as the WORC training and test datasets, respectively.

For external validation, we used the publicly available STT data of 55 patients released on The Cancer Imaging Archive (TCIA), which we refer to as the TCIA test dataset. **(Figure S1)** (14). For all patients, CT, T1-weighted, and T2-weighted-FS MRI are available. Reference segmentations were manually made by one expert radiation radiologist on the T2-weighted-FS MRI scans. The TCIA test dataset includes one modality (T2-weighted-FS MRI), and five tumor phenotypes (**Table 1**) which are not available in the WORC datasets.

### Interactive segmentation

Our interactive segmentation method is based on the framework by Luo et al., 2021 (12), as it has been designed for the medical domain, requires limited user interactions, and has been shown to generalize well to unseen objects. In our method (**Figure 1A**), users click six points near the extreme boundaries of the 3D object of interest, i.e. two extreme interior margin points in three planes (transversal, coronal, sagittal). Using these interior margin points the image is cropped to a region of interest (ROI) to aid the model in tumor localization. The cropping boundaries are slightly relaxed in order to encapsulate the whole object. The interior margin points are also used to calculate an exponentialized geodesic distance (EGD) map, which highlights the tumor voxels based on the intensity differences with the surrounding tissue. Together, the cropped image and EGD map are used as input for a 3D CNN model. Implementation details can be found in the **Supplementary Materials**.

### Fully automatic segmentation

To compare the performance of the interactive segmentation method, we trained a fully automatic segmentation method using the state-of-the-art self-configuring nnU-Net framework (**Supplementary Materials**) (9). nnU-Net deploys a 3D CNN on the whole volume in order to locate and segment the tumor. Prior to this work, nnU-Net has not yet been applied to STTs.

### Experimental setup

Separate segmentation models were trained for CT and T1-weighted MRI, both for the interactive and fully automatic approach. First, the models were validated on the WORC test dataset. Second, external validation, including assessment of generalization to unseen phenotypes and T2-weighted-FS MRI sequences not encountered during training, was performed on the TCIA test dataset. Here, the models trained on T1-weighted MRIs were used to generate segmentations for T2-weighted-FS MRI.

For both the WORC and TCIA dataset, the six interior margin points per image required for the interactive segmentation method were synthetically generated based on the reference segmentation. In order to validate synthetic interactions, a musculoskeletal radiologist (8 years of experience) and a medical student also performed real user interactions on the WORC test dataset. The model outputs generated from these real user interactions were then compared to the reference segmentation. Additional details can be found in the **Supplementary Materials**.

### Statistical analysis

Segmentation performance was evaluated using the Dice Similarity Coefficient (DSC) (15). To determine differences in DSC between automatic and interactive methods, we used a paired two-sided t-test.

Pyradiomics was used to calculate the volume and maximum diameter in the transverse plane of the predicted and the reference segmentation (16). Agreement between these measurements across segmentations was assessed using Bland-Altman plots and the Pearson correlation coefficient ($r$) (17).

### Data availability

To facilitate use of our interactive segmentation method, we provide a graphical user interface, a video tutorial, and the code for training and evaluation (18).

# Results

**Comparison on WORC test dataset of interactive, fully automatic, and reference segmentations**

Interactive segmentation resulted in a higher mean DSC and lower standard deviation (CT: 0.85±0.11, T1-weighted MRI: 0.84±0.12) than fully automatic segmentation (CT: 0.52±0.43, T1-weighted MRI: 0.71±0.35). Both differences were statistically significant (CT: p<0.001, T1-weighted MRI: p=0.007) (**Figure 2A, Table 2**).

Examples of interactive and fully automatic segmentations, including the synthetic interactions and EGD map, are shown in **Figure 1B**. Qualitative results, interobserver variability of two manual annotators, and the comparison of synthetic and real user interactions are described in the **Supplementary Materials**.

**External validation on TCIA test dataset including unseen tumor phenotypes and unseen imaging sequences**

The interactive segmentation method was able to segment unseen phenotypes on both CT and T1-weighted MRI with similar DSC to the phenotypes in the WORC test dataset (CT: 0.81±0.09, T1-weighted MRI: 0.84±0.09) (**Figure 2B, Table 2**). The automatic segmentation method performed similarly for unseen phenotypes on T1-weighted MRI (DSC: 0.81±0.23), however failed more often to segment tumors on CT (DSC: 0.38±0.34) compared to interactive segmentation (T1-weighted MRI: p=0.14, CT: p<0.001).

The detection of unseen phenotypes on the unseen modality, T2-weighted-FS MRI, improved slightly for the interactive segmentation method compared to T1-weighted MRI (DSC: 0.88±0.08). The automatic segmentation provided slightly worse results on T2-weighted-FS MRI (DSC:0.78±0.22) compared to the interactive segmentation (p=0.004). Better contrast between tumor and surrounding tissue on T2-weighted-FS in comparison to T1-weighted MRI may explain the slightly better segmentation results on T2-weighted-FS MRI (**Figure S2**).

**Agreement of volume and diameter measurements**

Volume and diameter measurements based on the interactive segmentation showed to have good agreement to the reference segmentation, both in the WORC test dataset (mean ± SD volume error: 1±28mm$^3$, r=0.99; diameter: -6±14mm, r=0.90) and in the TCIA test dataset (volume: -7±23mm$^3$, r=0.96; diameter: -3±6mm, r=0.99) **(Figure S3 and S4)**.

**Time-efficiency of segmentation methods**

The radiologist took on average 258 seconds (s) for CT and 122s for T1-weighted MRI to conduct interactive segmentation on the WORC dataset, which was considerably shorter than manual segmentation (CT: 1639s, T1-weighted MRI: 1895s) **(Table S2)** (5-7).

## Discussion

In this study, we developed a deep-learning method for minimally interactive segmentation of STTs on CT and MRI. Upon validation in the WORC test dataset, interactive segmentation achieved a high degree of overlap with the reference segmentation of clinicians for nine STT phenotypes (CT: 0.85±0.109, T1-weighted MRI: 0.84±0.121), with higher mean DSC and lower standard deviation compared to fully automatic segmentation. In the external TCIA test dataset, the interactive segmentation method showed good generalizability to the unseen phenotypes and MRI sequences (CT: 0.81±0.092, T1-weighted MRI: 0.84±0.092, T2-weighted-FS MRI: 0.88±0.075). In addition, there was good agreement for volume and diameter calculations between interactive and reference segmentations.

In the **Supplementary Materials** we motivate the need for time-efficient segmentation, discuss previous work on segmentation of STT, and point out additional considerations regarding clinical implementation of the minimally interactive segmentation method.

Our study has two main limitations. First, we evaluated our method on 14 STT phenotypes, while there are over 100 histological phenotypes. Although this study shows that our method generalized well to five unseen STT phenotypes, this is no guarantee that it also translates to other STT phenotypes, or even other cancer types beyond STT; hence, future work should investigate the further generalization to other tumor types. Second, either T1-weighted or T2-weighted-FS MRI was used as input to the method; a multimodal approach using different MRI sequences simultaneously may improve results further (19).

In conclusion, our minimally interactive deep learning-based segmentation method can accurately generate segmentations for a wide variety of STTs in different body parts imaged with CT or MRI. Therefore, this method could reduce the burden of manual segmentation for targeted (neo)adjuvant therapy, enable the integration of imaging-based biomarkers (e.g., radiomics) into clinical practice, and provide a fast, semi-automatic solution for volume and diameter measurements in the clinic.

**References:**

1. Sbaraglia M, Bellan E, Dei Tos AP. The 2020 WHO classification of soft tissue tumours: news and perspectives. Pathologica. 2021 Apr;113(2):70.

2. Gamboa AC, Gronchi A, Cardona K. Soft-tissue sarcoma in adults: an update on the current state of histiotype-specific management in an era of personalized medicine. CA: a cancer journal for clinicians. 2020 May;70(3):200-29.

3. Tiong SS, Dickie C, Haas RL, O'Sullivan B. The role of radiotherapy in the management of localized soft tissue sarcomas. Cancer biology & medicine. 2016 Sep;13(3):373.

4. Lambin P, Rios-Velazquez E, Leijenaar R, Carvalho S, Van Stiphout RG, Granton P, Zegers CM, Gillies R, Boellard R, Dekker A, Aerts HJ. Radiomics: extracting more information from medical images using advanced feature analysis. European journal of cancer. 2012 Mar 1;48(4):441-6.

5. Vos M, Starmans MP, Timbergen MJ, van der Voort SR, Padmos GA, Kessels W, Niessen WJ, Van Leenders GJ, Grünhagen DJ, Sleijfer S, Verhoef C. Radiomics approach to distinguish between well differentiated liposarcomas and lipomas on MRI. Journal of British Surgery. 2019 Dec;106(13):1800-9.

6. Timbergen MJ, Starmans MP, Padmos GA, Grünhagen DJ, van Leenders GJ, Hanff DF, Verhoef C, Niessen WJ, Sleijfer S, Klein S, Visser JJ. Differential diagnosis and mutation stratification of desmoid-type fibromatosis on MRI using radiomics. European Journal of Radiology. 2020 Oct 1;131:109266.

7. Starmans MP, Timbergen MJ, Vos M, Renckens M, Grünhagen DJ, van Leenders GJ, Dwarkasing RS, Willemssen FE, Niessen WJ, Verhoef C, Sleijfer S. Differential diagnosis and molecular stratification of gastrointestinal stromal tumors on CT images using a radiomics approach. Journal of Digital Imaging. 2022 Apr;35(2):127-36.

8. Ko CC, Yeh LR, Kuo YT, Chen JH. Imaging biomarkers for evaluating tumor response: RECIST and beyond. Biomarker research. 2021 Jul 2;9(1):52.

9. Isensee F, Jaeger PF, Kohl SA, Petersen J, Maier-Hein KH. nnU-Net: a self-configuring method for deep learning-based biomedical image segmentation. Nature methods. 2021 Feb;18(2):203-11.

10. Stiller CA, Trama A, Serraino D, Rossi S, Navarro C, Chirlaque MD, Casali PG, RARECARE Working Group. Descriptive epidemiology of sarcomas in Europe: report from the RARECARE project. European journal of cancer. 2013 Feb 1;49(3):684-95.

11. Stacchiotti S, Frezza AM, Blay JY, Baldini EH, Bonvalot S, Bovée JV, Callegaro D, Casali PG, Chiang RC, Demetri GD, Demicco EG. Ultra-rare sarcomas: A consensus paper from the Connective Tissue Oncology Society community of experts on the incidence threshold and the list of entities. Cancer. 2021 Aug 15;127(16):2934-42.

12. Luo X, Wang G, Song T, Zhang J, Aertsen M, Deprest J, Ourselin S, Vercauteren T, Zhang S. MIDeepSeg: Minimally interactive segmentation of unseen objects from medical images using deep learning. Medical image analysis. 2021 Aug 1;72:102102.

13. Starmans MP, Timbergen MJ, Vos M, Padmos GA, Grünhagen DJ, Verhoef C, Sleijfer S, van Leenders GJ, Buisman FE, Willemssen FE, Koerkamp BG. The WORC database: MRI and CT scans, segmentations, and clinical labels for 930 patients from six radiomics studies. medRxiv. 2021 Aug 25:2021-08.

14. Vallières M, Freeman CR, Skamene SR, El Naqa I. A radiomics model from joint FDG-PET and MRI texture features for the prediction of lung metastases in soft-tissue sarcomas of the extremities. Physics in Medicine & Biology. 2015 Jun 29;60(14):5471.

15. Zou KH, Warfield SK, Bharatha A, Tempany CM, Kaus MR, Haker SJ, Wells III WM, Jolesz FA, Kikinis R. Statistical validation of image segmentation quality based on a spatial overlap index1: scientific reports. Academic radiology. 2004 Feb 1;11(2):178-89.

16. Van Griethuysen JJ, Fedorov A, Parmar C, Hosny A, Aucoin N, Narayan V, Beets-Tan RG, Fillion-Robin JC, Pieper S, Aerts HJ. Computational radiomics system to decode the radiographic phenotype. Cancer research. 2017 Nov 1;77(21):e104-7.

17. Altman DG, Bland JM. Measurement in medicine: the analysis of method comparison studies. Journal of the Royal Statistical Society Series D: The Statistician. 1983 Sep;32(3):307-17.

18. Douwe J. Spaanderman, Martijn P. A. Starmans, Stefan Klein. Douwe-Spaanderman/InteractiveNet: v0.2.1. Zenodo; 2023.

19. Guo Z, Li X, Huang H, Guo N, Li Q. Deep learning-based image segmentation on multimodal medical imaging. IEEE Transactions on Radiation and Plasma Medical Sciences. 2019 Jan 1;3(2):162-9.

# Tables

**Table 1: Descriptive characteristics of the datasets used in this study.**

| Characteristics | WORC training dataset (n = 412) [13] | WORC test dataset (n = 102) [13] | TCIA test dataset (n = 51) [14] |
|---|---|---|---|
| Age (y) | | | |
|    Range | 5-93 | 1-86 | 16–83 |
|    Mean ± SD | 57 ± 17 | 55 ± 20 | 55 ± 17 |
| Sex | | | |
|    Female | 206 (50) | 54 (53) | 27 (53) |
|    Male | 206 (50) | 48 (47) | 24 (47) |
| Location | | | |
|    Lower extremities | 156 (38) | 34 (33) | 47 (92) |
|    Upper extremities | 27 (7) | 7 (7) | 4 (8) |
|    Head and neck | 14 (3) | 4 (4) | - |
|    Intra-abdominal | 159 (9) | 39 (38) | - |
|    Retroperitoneum and pelvis | 3 (1) | - | - |
|    Trunk | 53 (13) | 18 (18) | - |
| Volume (mm$^3$) | | | |
|    Range | 0.6-7944 | 1.1-3138 | 17-2361 |
|    Mean ± SD | 379 ± 837 | 305 ± 505 | 474 ± 499 |
| Modality | | | |
|    CT | 158 (38) | 39 (38) | 51 (100) |
|    T1-weighted MRI | 254 (62) | 63 (62) | 51 (100) |
|    T2-weighted fat suppressed MRI | - | - | 51 (100) |
| Phenotype | | | |
|    Lipoma | 46 (11) | 11 (11) | - |
|    Well-differentiated liposarcoma | 46 (11) | 11 (11) | - |
|    Desmoid-type fibromatosis | 58 (14) | 14 (14) | - |
|    Myxofibrosarcoma | 49 (12) | 12 (12) | - |
|    Myxoid liposarcoma | 29 (7) | 8 (8) | - |
|    Gastro-intestinal stromal tumor | 100 (24) | 24 (24) | - |
|    Schwannoma | 18 (4) | 5 (5) | - |
|    Leiomyosarcoma | 46 (11) | 12 (12) | 10 (20) |
|    Leiomyoma | 20 (5) | 5 (5) | - |
|    Liposarcoma | - | - | 11 (21) |
|    Fibrosarcoma | - | - | 1 (2) |
|    Synovial sarcoma | - | - | 5 (10) |
|    Malignant fibrous histiocytoma | - | - | 17 (33) |
|    Extraskeletal bone sarcoma | - | - | 4 (8) |
|    Other | - | - | 3 (6) |

Except where indicated, data are the number of patients (percentages).

**Table 2: Phenotype-specific agreement between the fully automatic or interactive segmentation method and reference segmentation.**

|  | CT | | T1-weighted MRI | | T2-weighted FS MRI | |
|---|---|---|---|---|---|---|
|  | automatic | interactive | automatic | interactive | automatic | interactive |
| **WORC test dataset** | | | | | | |
|   GIST | 0.49±0.43* | 0.84±0.12* | | | | |
|   Schwannoma | 0.73±0.37 | 0.87±0.09 | | | | |
|   Leiomyoma | 0.28±0.35 | 0.84±0.01 | | | | |
|   Leiomyosarcoma | 0.71±0.36 | 0.86±0.08 | 0.88±0.05 | 0.88±0.04 | | |
|   Lipoma | | | 0.73±0.35 | 0.90±0.05 | | |
|   MLS | | | 0.72±0.36 | 0.84±0.13 | | |
|   DTF | | | 0.49±0.39 | 0.75±0.17 | | |
|   Myxofibrosarcoma | | | 0.69±0.33 | 0.82±0.11 | | |
|   WDLS | | | 0.86±0.27 | 0.89±0.06 | | |
|     *Total* | 0.52±0.43* | 0.85±0.11* | 0.71±0.35* | 0.84±0.12* | | |
| **TCIA test dataset** | | | | | | |
|   Liposarcoma | 0.40 ± 0.30* | 0.82 ± 0.09* | 0.83 ± 0.22 | 0.84 ± 0.10 | 0.84 ± 0.18 | 0.90 ± 0.06 |
|   Leiomyosarcoma | 0.49 ± 0.36* | 0.87 ± 0.03* | 0.82 ± 0.27 | 0.87 ± 0.07 | 0.81 ± 0.20 | 0.88 ± 0.07 |
|   Fibrosarcoma[§] | 0.31 | 0.79 | 0.78 | 0.59 | 0.94 | 0.88 |
|   Synovial sarcoma | 0.18 ± 0.36* | 0.84 ± 0.04* | 0.79 ± 0.20 | 0.87 ± 0.06 | 0.75 ± 0.24 | 0.90 ± 0.05 |
|   MFH | 0.39 ± 0.33* | 0.77 ± 0.07* | 0.77 ± 0.20 | 0.83 ± 0.10 | 0.72 ± 0.23 | 0.86 ± 0.09 |
|   ESBS | 0.40 ± 0.40 | 0.75 ± 0.13 | 0.65 ± 0.37 | 0.83 ± 0.07 | 0.78 ± 0.18 | 0.83 ± 0.06 |
|   Other | 0.29 ± 0.23 | 0.84 ± 0.02 | 0.89 ± 0.02 | 0.90 ± 0.03 | 0.76 ± 0.24 | 0.90 ± 0.03 |
|     *Total* | 0.38 ± 0.34* | 0.81 ± 0.08* | 0.79 ± 0.23 | 0.84 ± 0.09 | 0.78 ± 0.22* | 0.88 ± 0.08* |

Data are mean ± SD for the Dice Similarity Coefficient (DSC). FS: fat saturated; MLS: myxoid-liposarcoma; DTF: desmoid-type fibromatosis; WDLS: well-differentiated liposarcoma; GIST: gastrointestinal stromal tumor; MFH: malignant fibrous histiocytoma; and ESBS: Extraskeletal bone sarcoma.

[§] Standard deviation could not be calculated as only one fibrosarcoma was present in the TCIA dataset.

* P < .05 (paired t-test of DSC between interactive and automatic segmentation).

# Figures

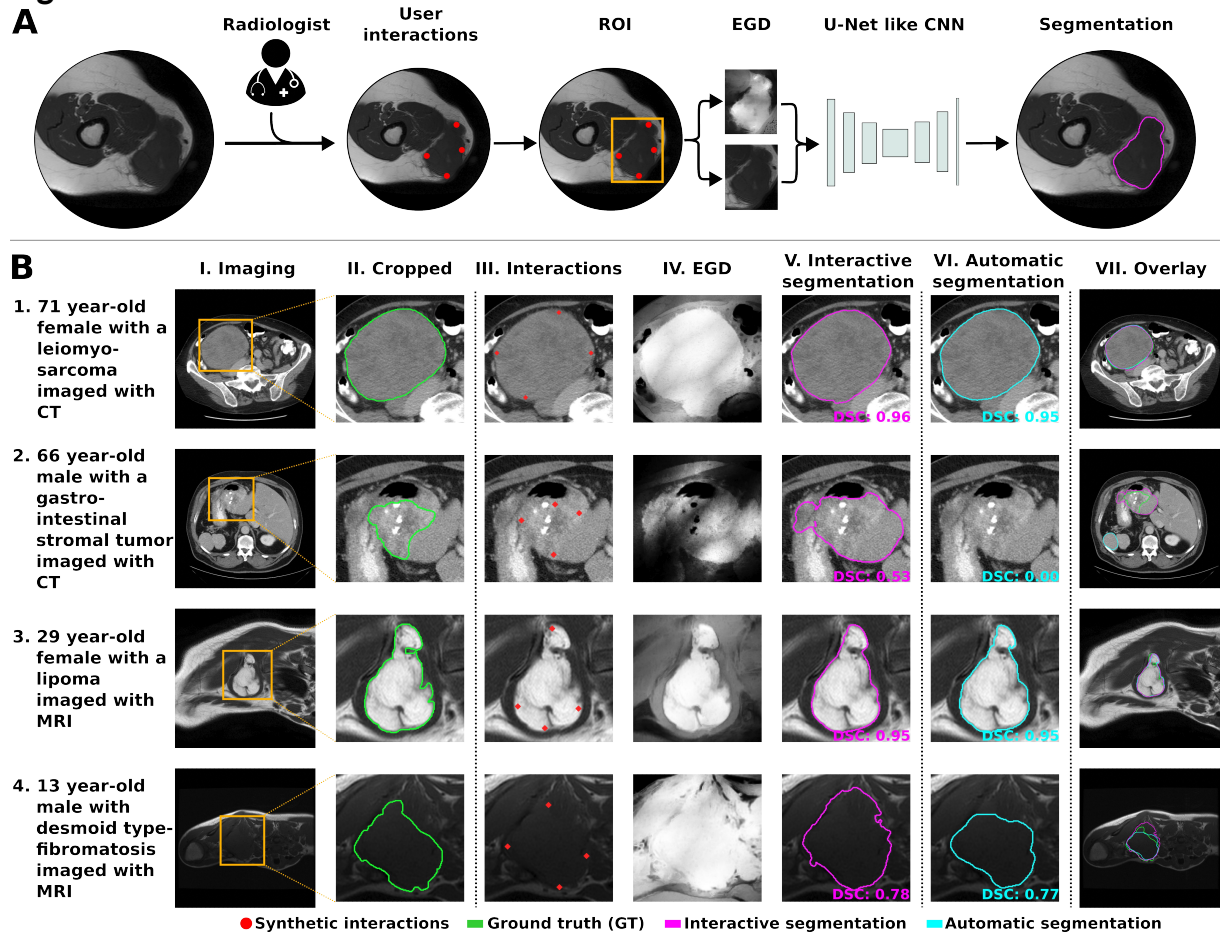

**Figure 1**: **A)** Schematic overview of the interactive framework, based on (12), used in this study. For visualization purposes, one 2D slice in the transverse plane is shown, while all data are 3D images and all operations are performed in 3D. In the interactive segmentation pipeline, a clinician has to draw six interior margin points in the 3D object. Next, using these interactions the region of interest (ROI) is extracted, and the exponentialized geodesic distance map (EGD) is calculated. The cropped image and the EGD map are concatenated and fed through a U-Net-like Convolutional Neural Network (CNN). **B)** Qualitative comparison of interactive and fully automatic segmentation methods in four different patients with STT in the WORC test dataset; The images show **I)** CT imaging (rows 1 and 2) or T1-weighted MRI (rows 3 and 4) in the transverse plane, **II)** zoomed-in image with reference segmentation for visualization purposes, **III)** interior margin points derived synthetically, **IV)** EGD map derived from the interior margin points, **V)** predicted interactive segmentation, **VI)** predicted fully automatic segmentation, **VII)** segmentation comparison in full image.

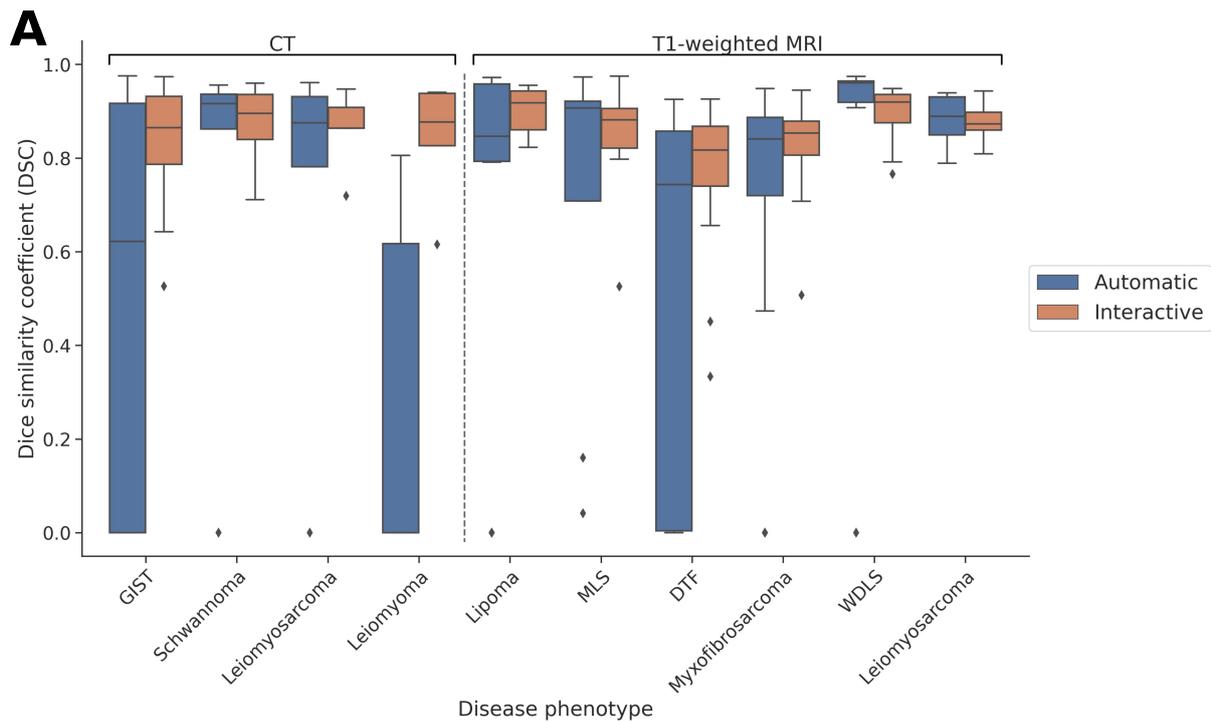

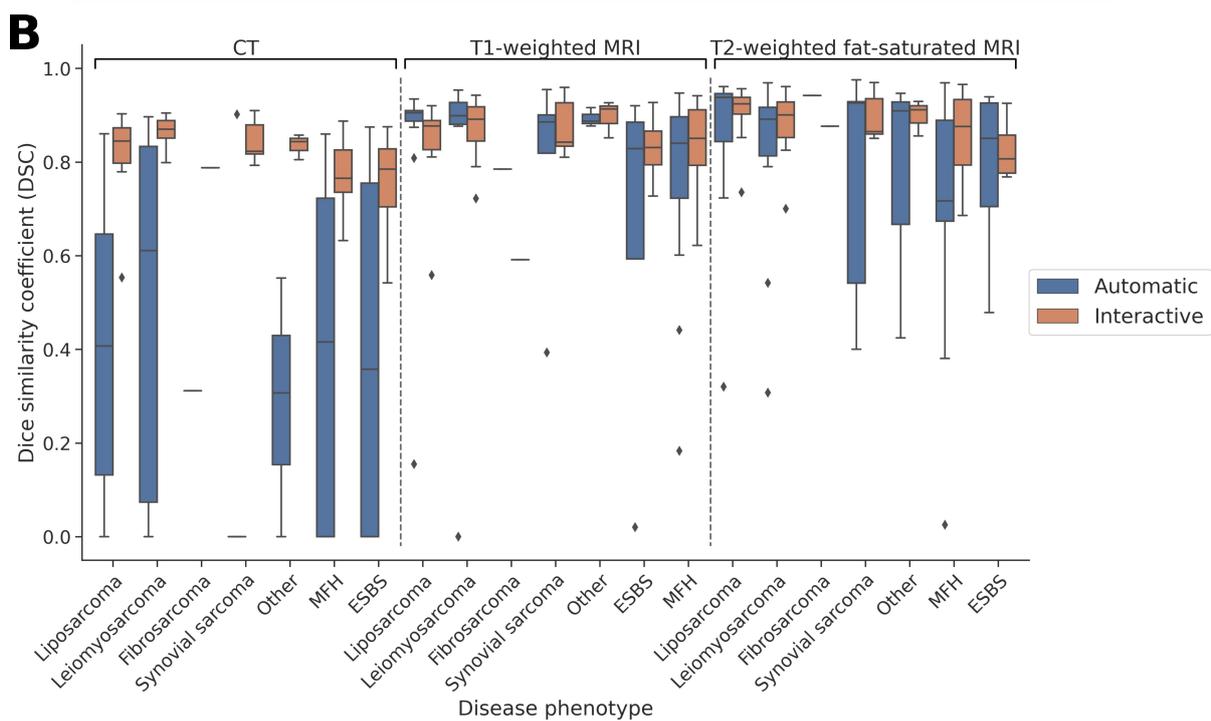

**Figure 2**: Quantitative results from automatic and interactive segmentation of STTs on the **A)** WORC test dataset, and **B)** TCIA test dataset. Box-and-whisker plot, visualizing the median, quartiles, and potential outliers, for the Dice Similarity Coefficient (DSC) results of fully automatic (blue) and interactive (orange) segmentation methods for different phenotypes on CT and MRI. Higher DSC means better segmentation accuracy. MLS: myxoid-liposarcoma; DTF: desmoid-type fibromatosis; WDLS: well-differentiated liposarcoma; GIST: gastrointestinal stromal tumor; MFH: malignant fibrous histiocytoma; and ESBS: Extraskeletal bone sarcoma.

# Supplementary Material
## Materials and Methods
### Interactive segmentation

For the preprocessing, network architecture, training, and post-processing, we follow most of the best practices found in the nnU-Net pipeline (9). The design choices of our interactive segmentation implementation are divided into 1) fixed parameters, which are predefined and do not change based on the application; 2) rule-based parameters, which are based on the characteristics of the dataset; 3) empirical parameters, which are determined based on achieved results.

We provide the code and model weights for training and evaluation our self-configuring interactive segmentation method (18), including a graphical user interface, implemented in MONAI Label, in which the clinician can draw the six interior margin points and run the interactive segmentation pipeline (20).

The segmentation methods were implemented in Python (3.9.5) and Pytorch (1.21.1) using a NVIDIA A40 (48GB) GPU.

### Fixed parameters

*Region of interest selection*. We use the same strategy as Luo et al, 2021 (12), where region of interest (ROI) extraction is based on the interior margin points from the user. In order to make sure that the complete object is incorporated in the image and the dynamics at the boundary of the object is fully captured, the ROI is expanded slightly by a relaxation factor of 0.1 times the size of the ROI in each axis. Finally, if the input is not divisible by 2 to the power of the number of downsampling operations, the ROI is further extended, possibly using zero-padding in the case that the ROI falls outside the image.

*Deep learning model architecture*. We use a convolutional neural network (CNN), more specifically a 3D U-Net-like network with instance normalization, leaky ReLU, and deep supervision similar to nnU-Net (9). Down- and upsampling are implemented as respectively a stride and transpose convolution. In line with the original work from Luo et al., 2021 (12), as the images are cropped to the ROI and are therefore already much smaller than the original image, we reduce the number of feature maps to start at 4 and double/halve with each down-/upsampling operation, in order to limit the required memory consumption, and achieve real-time inference.

*Training*. We use five-fold cross-validation to train five CNNs. These models are trained for 1000 epochs, using a poly learning rate scheduler (initial learning rate = 0.01, learning rate decay = $(1-epoch/epoch_{max})^{0.9}$), Dice + cross-entropy loss, stochastic gradient descent with Nesterov momentum ($\mu$ = 0.99), and deep supervision with additional auxiliary losses in the decoder for all but the lowest two resolution. During training, we randomly augment inputs using rotations, zooms, and flips. Additionally, the image is randomly altered using Gaussian noise, Gaussian smoothing, intensity scaling, and contrast adjustment. These configurations are all in line with nnU-Net. However, as we are dealing with unequal-sized images, we cannot create batch sizes larger than 1. To stabilize training, we use gradient accumulation (n = 4), i.e. updating the weights after four iterations.

*Inference*. We use all five models during inference time, and combine the results with mean ensembling. Also, for each model we use test time augmentation, where we flip the image in all directions, e.g. eight different flips for 3D images, and combine the results with mean ensembling.

**Rule-based parameters**

The rule-based parameters are based on the characteristics of the training dataset. For prediction on new samples, we use the same rule-based parameters as used during training.

*Resampling*. The physical space a voxel represents (spacing) is often heterogeneous between different medical images, which needs to be homogenized before providing the image to the CNN. Here we use the nnU-Net resampling method, in which the median spacing for each axis in the training dataset is used as target spacing. In the case of anisotropic (maximum axis spacing / minimum axis spacing > 3) resampling, which is the case for the WORC dataset, the tenth percentile is taken for the lowest resolution axis. This is done to conserve as much resolution in this axis as possible. Images are resampled with third-order spline and anisotropic out-of-plane dimension with nearest neighbor. Similarly, segmentations are resampled with linear interpolation and out-of-plane with nearest neighbor. The interactions from the clinician also have to be resampled to the new voxel spacing. Therefore, we first change the interactions to real-world coordinates and subsequently map these coordinates to the location in the resized image.

*Normalization*. We use the same normalization as deployed by nnU-Net, in which for CT, normalization is done by percentile clipping and z-scoring based on the voxels in the tumor (foreground voxels) in the training dataset. For MRI, normalization is done per image using z-scoring. In contrast to nnU-Net, normalization for both modalities is done based on the extracted ROI only.

*Network topology*. In line with nnU-Net, by default a 3 x 3 x 3 kernel size for convolution is used. However, when dealing with anisotropic images, pseudo-3D kernels, e.g. 3 x 3 x 1, can be deployed to deal with resolution discrepancy. Similar to nnU-Net, resolution discrepancy is defined by a spacing ratio larger than two. Given the anistotropic images in the WORC dataset, the initial two convolution operations are done using this 3 x 3 x 1 convolution.

**Empirical parameters**

*Post-processing*. Post-processing involves the utilization of either the largest connected component, filling small holes, or both. We assess the effect of post-processing on improving the DSC of the segmentation compared to the reference segmentation. We evaluate this by comparing the DSC in cross-validation splits both with and without post-processing methods. For CT and MRI data in the WORC dataset, both these post-processing steps turned out to be effective, and were thus employed.

**Synthetic interactions**

The interactive segmentation method requires six interior margin points per image. For both the WORC and TCIA dataset, these six points were generated synthetically based on the reference segmentation. First, extreme points were identified along all axes using the reference segmentation. Next, to make sure the points were inside the object, these points were moved five voxels inwards in-plane, and in case of anisotropic images (maximum axis spacing / minimum axis spacing > 3), in the out-of-plane dimension by 1 voxel.

**Real user interactions, time measurement and quality scoring**
To validate the synthetic interactions, real user interactions were performed on the WORC test dataset by a musculoskeletal radiologist with 8 years of experience, and an untrained medical student. They were only provided the images and blinded to clinical data, including STT phenotype. The users had not seen these images prior to annotation and the reference segmentation was not shown during annotation. We evaluated: 1) performance of real user versus synthetic interactions; 2) expertise required to draw interactions; 3) user-provided quality scores based on four scales (Excellent, Sufficient, Insufficient, Incorrect, see **Table S3** for details); and 4) time to draw the interactions, run the pipeline, and score segmentation quality.

## Results
### Qualitative evaluation on WORC test dataset
Qualitative analysis showed that poor contrast between the tumor and surrounding tissue sometimes led to low DSC segmentation using either method (**Figure 1B**). Additionally, both methods had difficulties with tumor boundaries when tumors were irregular and lobulated. Fully automatic segmentation had difficulties with large fields of view, especially in CT, as tumor detection showed to be difficult, sometimes resulting in no segmentation (CT: n=5/39, T1-weighted MRI: n=5/62), or segmentation of a different object (CT: n=10/39, T1-weighted MRI: n=3/62). Nevertheless, even after removing the 23 outliers on which the automatic method failed to segment the correct lesion, the interactive method outperformed the fully automatic method in terms of DSC, with a higher overall median, smaller IQR, and fewer outliers (**Figure S5**).

### Interobserver variability between manual annotators
For a subset of the WORC dataset (n=60), the interobserver DSC was reported between two manual annotators (CT: 0.84±0.20, T1-weighted MRI: 0.77±0.20) (5-7). This shows that the performance of the interactive segmentation matched or exceeded intra-observer variability (CT: 0.85±0.11, T1-weighted MRI: 0.84±0.12). Moreover, the standard deviation of the interactive segmentation was lower against the reference segmentation, compared to the two manual annotators, suggesting that interactive segmentation reduces variability among produced segmentations.

### Comparison on WORC test dataset of synthetic and real user interactions
We compared the interactive segmentation method with synthetic and real user interactions on the WORC test dataset **(Figure S6, Table S4)**. The musculoskeletal radiologist's interactions performed significantly worse on the T1-weighted MRI data (DSC: 0.75 ± 0.25) in comparison to the synthetic interactions (DSC: 0.84 ± 0.12, p=0.01), whereas the medical student scored significantly worse on the CT data (DSC: 0.60 ± 0.40) in comparison to the synthetic interactions (DSC: 0.85 ± 0.11, p=0.001). The difference in DSC between the two users and the synthetic interactions was mostly explained by the number of wrongly identified objects (radiologist: n=7/102, student: n=16/102). Excluding these samples increased the DSC for both the radiologist (CT: 0.81±0.15, T1-weighted MRI: 0.79±0.19) and medical student (CT: 0.84±0.14, T1-weighted MRI: 0.84±0.15). After excluding these samples, significant differences remained only for T1-weighted MRI for the radiologist (T1-weighted MRI: p=0.03). Qualitative scoring from the radiologist showed that most segmentations were deemed Sufficient or Excellent (CT: n=30/39, T1-weighted MRI: n=42/63), see **Table S5**.

## Discussion
### The need for time-efficient segmentation
The ability to provide time-efficient segmentation on CT and MRI is important for the translation of morphological, volumetric quantifications, and radiomics biomarkers to clinical practice.

For example, segmentations are required for targeted (neo)adjuvant radiotherapy (3). Currently, these have to be made manually for each lesion at each regiment, which is a substantial burden on the physician's time and drives healthcare costs. Also, the use of quantitative imaging features (i.e., radiomics) has often been described to provide accurate, noninvasive imaging biomarkers (4). In STT, radiomics has been used to predict phenotype (5-7) grading (21-23), and patient outcome (23-25). However, a major hurdle for translation of radiomics to clinical practice is the requirement for manual segmentation, which is observer dependent, time-consuming and therefore not feasible in the radiologist's workflow. Recent work by Crombé et al., 2020, and Gitto et al., 2021, showed that none of the 52 and 49 included radiomics studies, respectively, were performed with automatic segmentation (26, 27). Only 13.5% and 8.0% used semi-automatic methods, and unlike the minimally interactive approach provided here, these methods still required intensive manual correction by the clinician. Finally, routine clinical measurements, such as the Response Evaluation Criteria In Solid Tumors (RECIST), would benefit from segmentation to reduce observer variability, or allow extension to 3D volume measurements, which has been shown to improve monitoring of therapy response (8, 28). Together, these examples highlight the need for (semi-)automatic segmentation in clinical practice.

Further research should explore the use of minimally interactive segmentation for these applications including the impact on the clinical workflow and decision making. Future research also includes extension of our method to other cancer types beyond STT. Due to the wide variety of STT phenotypes, locations, and imaging appearance, we expect that our method may even work directly on other (similar) tumors without any adjustments.

### Prior work on segmentation methods for STT
Some prior work has focused on deep-learning-based segmentation for STT. First, a study using the TCIA database showed accurate results (DSC=0.88) when combining FDG-PET, CT, and T2-weighted-FS MRI for automatic segmentation (19). However, this study only assessed STT in the extremities, lacked external validation, and required all three modalities to provide accurate segmentations. Next, recent work achieved accurate automatic segmentation of lipomatous tumors (0.80 ± 0.184) but lacked external validation and focused on a single STT type (29). Here, we provide an interactive segmentation method with improved segmentation results that generalizes well to different phenotypes of STT.

## Considerations regarding clinical implementation

Here, we delve into the practical considerations for implementing our interactive segmentation method in clinical practice. Specifically, we address real user interactions, the implications of quality scoring, and the role of fully automatic segmentation.

In this study, we derived synthetic interactions from the reference segmentations to train and validate our method, since using real interactions would have been time-consuming due to the large number of patients. To verify that synthetic interactions reflected those of humans, we assessed both types on the WORC test dataset. We found differences between real and synthetic interactions for the radiologist and medical student. However, this was mostly explained by segmentation of the wrong object by the user, as removing these samples yielded better results for both users. Furthermore, it is worth noting that in real clinical workflows, clinicians have access to additional information and multiple MRI sequences, reducing the likelihood of tumor misidentification compared to the setting used in our study with only the T1-weighted MRI data. Further research should explore the performance of the interactive segmentation method within real clinical settings to assess its practical impact.

Next, aside from comparing interactive segmentations to reference segmentations using DSC, interactive segmentation based on real interactions were also quality scored by the user. Both users scored most segmentations as Excellent or Sufficient. Future research should investigate the implications of these quality scores for clinical measurements; e.g. targeted radiotherapy might require Excellent segmentation while radiomics might perform adequately on Sufficient segmentations. Furthermore, still a substantial number of segmentations were scored as Insufficient, suggesting that manual correction may still be necessary in these cases. Nevertheless, interactive segmentation, where radiologists are already involved, may align more naturally with clinical workflows compared to fully automatic segmentation methods when adjustments are required.

Finally, the primary focus is on developing and implementing interactive segmentation. However, we also developed a fully automatic segmentation method for STT. While interactive segmentation generally outperformed fully automatic segmentation, the fully automatic method can still be valuable for certain patients, such as those with lipomas or leiomyosarcomas. Nevertheless, with fully automatic segmentation, the user should be wary for segmentation of the wrong object (CT: n=10/39, T1-weighted MRI: n=3/62). Therefore, expert knowledge is required regardless of the segmentation method applied.

**Reference**

20.  Diaz-Pinto A, Alle S, Ihsani A, Asad M, Nath V, Pérez-Garcia F, et al. MONAI Label: A framework for AI-assisted Interactive Labeling of 3D Medical Images. arXiv preprint arXiv:2203.12362. 2022 Mar 23.

21.  Peeken JC, Spraker MB, Knebel C, Dapper H, Pfeiffer D, Devecka M, et al. Tumor grading of soft tissue sarcomas using MRI-based radiomics. EBioMedicine. 2019;48:332–40.

22.  Zhang Y, Zhu Y, Shi X, Tao J, Cui J, Dai Y, et al. Soft tissue sarcomas: preoperative predictive histopathological grading based on radiomics of MRI. Acad Radiol. 2019;26(9):1262–8.

23.  Peeken JC, Bernhofer M, Spraker MB, Pfeiffer D, Devecka M, Thamer A, et al. CT-based radiomic features predict tumor grading and have prognostic value in patients with soft tissue sarcomas treated with neoadjuvant radiation therapy. Radiother Oncol. 2019;135:187–96.

24.  Spraker MB, Wootton LS, Hippe DS, Ball KC, Peeken JC, Macomber MW, et al. MRI radiomic features are independently associated with overall survival in soft tissue sarcoma. Adv Radiat Oncol. 2019;4(2):413–21.

25.  Crombé A, Le Loarer F, Sitbon M, Italiano A, Stoeckle E, Buy X, et al. Can radiomics improve the prediction of metastatic relapse of myxoid/round cell liposarcomas? Eur Radiol. 2020;30(5):2413–24.

26.  Crombé A, Fadli D, Italiano A, Saut O, Buy X, Kind M. Systematic review of sarcomas radiomics studies: Bridging the gap between concepts and clinical applications? Eur J Radiol. 2020 Nov;132:109283.

27.  Gitto S, Cuocolo R, Albano D, Morelli F, Pescatori LC, Messina C, et al. CT and MRI radiomics of bone and soft-tissue sarcomas: a systematic review of reproducibility and validation strategies. Insights Imaging. 2021 Jun;12(1).

28.  Mozley PD, Bendtsen C, Zhao B, et al. Measurement of tumor volumes improves RECIST-based response assessments in advanced lung cancer. Transl Oncol. 2012;5(1):19-25.

29.  Liu CC, Abdelhafez YG, Yap SP, et al. AI-based automated lipomatous tumor segmentation in MR images: ensemble solution to heterogeneous data. J Digit Imaging. 2023;36(3):1049-1059.

# Supplementary Tables

## Supplementary Table S1: Properties of the acquisition protocols

| Protocol | WORC training dataset (n = 412) [15] | | WORC test dataset (n = 102) [15] | | TCIA test dataset (n = 51) [16] | |
|---|---|---|---|---|---|---|
| Sequence | T1-weighted MRI (n=254) | CT (n=158) | T1-weighted MRI (n=63) | CT (n=39) | T1- and T2-weighted-FS MRI (n=51) | CT (n=51) |
| Manufacturer | | | | | | |
|   Siemens | 110 | 71 | 27 | 19 | 8 | |
|   Philips | 102 | 48 | 25 | 10 | 8 | |
|   General Electric | 42 | 7 | 11 | | 33 | 51 |
|   Canon | | 29 | | 7 | | |
|   Toshiba | | 3 | | 3 | | |
|   Varian | | | | | 2 | |
| Magnetic field strength | | | | | | |
|   1T | 22 | - | 8 | - | NR | - |
|   1.5T | 211 | - | 52 | - | NR | - |
|   3T | 21 | - | 3 | - | NR | - |
| Slice thickness (mm)* | 4.67 ± 1.27 | 4.21 ± 1.17 | 4.77 ± 1.61 | 4.20 ± 1.09 | 5.70 ± 1.32 | 3.80 ± 0.00 |
| Pixel spacing (mm)* | 0.67 ± 0.25 | 0.75 ± 0.09 | 0.72 ± 0.33 | 0.72 ± 0.07 | 0.77 ± 0.32 | 0.98 ± 0.00 |

Note. – Abbreviations: T: tesla, NR: Not reported. FS: fat saturated.

* Data are mean ± SDs in seconds.

**Supplementary Table S2: Time to perform minimally interactive segmentation, manual segmentation and fully automatic segmentation.**

|  | CT | | | | T1-weighted MRI | | | |
|---|---|---|---|---|---|---|---|---|
|  | Radiologist | Medical student | Manual* | Fully automatic† | Radiologist | Medical student | Manual* | Fully automatic† |
| Annotation | 146 ± 103 | 247 ± 608 | | | 87 ± 99 | 210 ± 538 | | |
| Preprocessing | 21 ± 12 | 21 ± 12 | | | 2.8 ± 2.2 | 2.7 ± 2.4 | | |
| Model inference | 1.3 ± 1.8 | 1.3 ± 1.8 | | | 0.4 ± 0.5 | 0.4 ± 0.4 | | |
| Postprocessing | 21 ± 10 | 17 ± 7.6 | | | 2.0 ± 1.6 | 2.0 ± 1.7 | | |
| Evaluation | 70 ± 84 | 42 ± 37 | | | 27 ± 23 | 160 ± 774 | | |
| User total | 216 ± 135 | 325 ± 617 | | | 117 ± 106 | 383 ± 840 | | |
| Method total | 43 ± 19 | 39 ± 17 | | 1940 ± 1113 | 5.2 ± 3.7 | 5.1 ± 4.0 | | 100 ± 95 |
| Total | 258 ± 135 | 364 ± 626 | 1895 ± 1804 | | 122 ± 107 | 388 ± 840 | 1639 ± 1397 | |

Note. – Data are mean ± SDs in seconds. Results are reported on the WORC test dataset.

User total: annotation + evaluation; Method total: preprocessing + model inference + postprocessing.

* Time to create the manual reference segmentation is also reported for comparison.

† Total inference time of the fully automatic segmentation method is also reported for comparison.

**Supplementary Table S3: Rule-based scoring of tumor segmentation based on visual inspection by the user**

| Segmentation score | Definition |
| --- | --- |
| Excellent | The segmentation is perfectly aligned with the tumor and requires no adjustments. For this score, the segmentation volume should overlap with the tumor for at least 95%. |
| Sufficient | The segmentation is aligned with the tumor, however, could benefit from minor adjustments. For this score, the segmentation volume should overlap with the tumor for at least 75%. |
| Insufficient | The segmentation misses parts of the tumor, or parts are overlapping with normal tissue, therefore major adjustments are required. For this score, the segmentation volume should overlap with the tumor for at least 50%. |
| Incorrect | The segmentation is not overlapping with the tumor, or missing large areas of the tumor. For this score, the segmentation volume does not overlap with the tumor for <50%. |

Note. – If the tumor cannot be located in the image, the clinician can score as "Cannot locate tumor".

**Supplementary Table S4: Comparison of interactive segmentation performance using interactions by different annotators.**

|  | Radiologist | Medical student | Synthetic | P value for synthetic vs radiologist* | P value for synthetic vs student* | P value for radiologist vs student* |
|---|---|---|---|---|---|---|
| CT |  |  |  |  |  |  |
|   All tumors | 0.77 ± 0.24 | 0.60 ± 0.40 | 0.85 ± 0.11 | .10 | .001 | .004 |
|   Correctly identified tumors† | 0.81 ± 0.15 | 0.84 ± 0.14 |  | .07 | .12 | .78 |
| T1-weighted MRI |  |  |  |  |  |  |
|   All tumors | 0.75 ± 0.25 | 0.77 ± 0.27 | 0.84 ± 0.12 | .01 | .07 | .15t |
|   Correctly identified tumors† | 0.79 ± 0.19 | 0.84 ± 0.15 |  | .03 | .36 | .02 |

Note. – Except where indicated, data are mean ± SDs for the Dice Similarity Coefficient (DSC). Results are reported on the WORC test dataset.

* P-values are reported for the paired t-test.

† Number of correctly identified tumors differ between radiologist and student.

**Supplementary Table S5: Quality scoring determined through visual inspection of the interactive segmentation made using interactions by different annotators.**

|  | CT | | T1-weighted MRI | |
|---|---|---|---|---|
|  | Radiologist | Medical Student | Radiologist | Medical Student |
| Excellent | 18 | 4 | 16 | 10 |
| Sufficient | 12 | 22 | 26 | 30 |
| Insufficient | 7 | 7 | 16 | 15 |
| Incorrect | 2 | 3 | 3 | 4 |
| Cannot locate tumor | 0 | 3 | 1 | 3 |

Note. – Results are reported on the WORC test dataset.

# Supplementary Figures

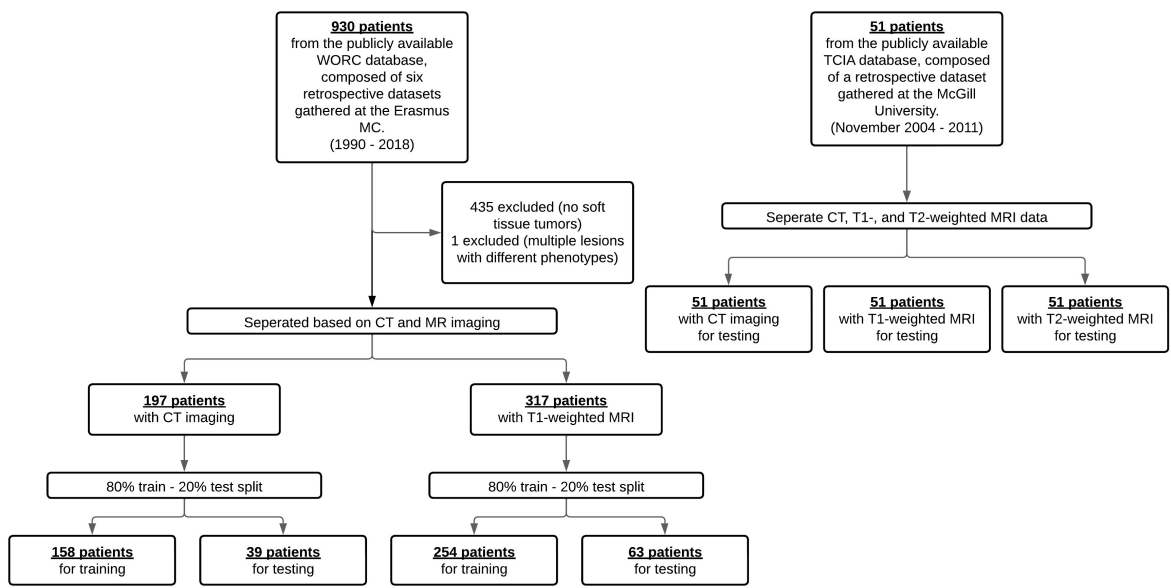

**Supplementary Figure S1**: Flowchart showing the two publicly available, retrospective, datasets used in this study and the exclusion criteria.

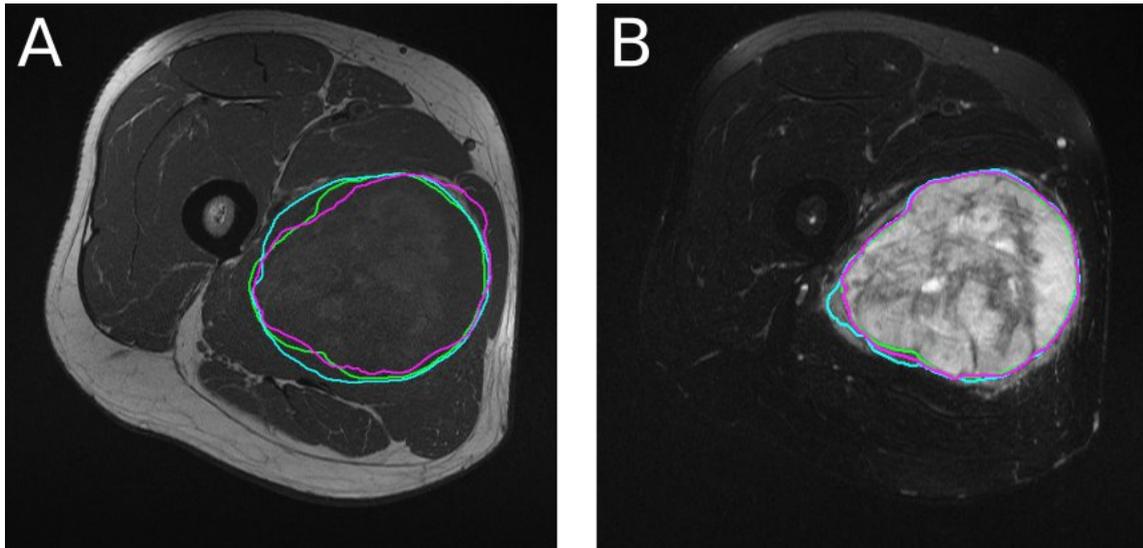

**Supplementary Figure S2:** Example from the TCIA test dataset with segmentation methods on **A)** T1-weighted MRI in the transverse plane, **B)** T2-weighted fat saturated MRI in the transverse plane. The reference segmentation is displayed in green, interactive segmentation in pink, and automatic segmentation in cyan. The contrast in the T2-weighted fat saturated image between the tumor and surrounding tissue improved both segmentation results.

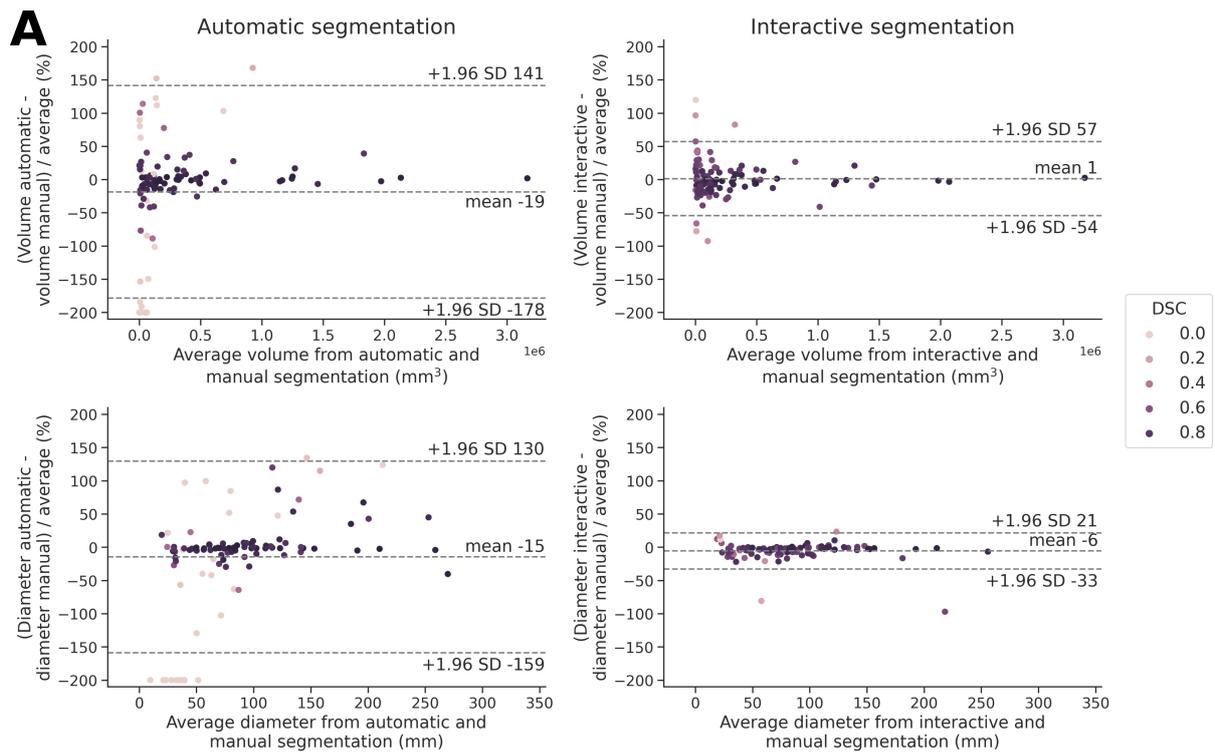
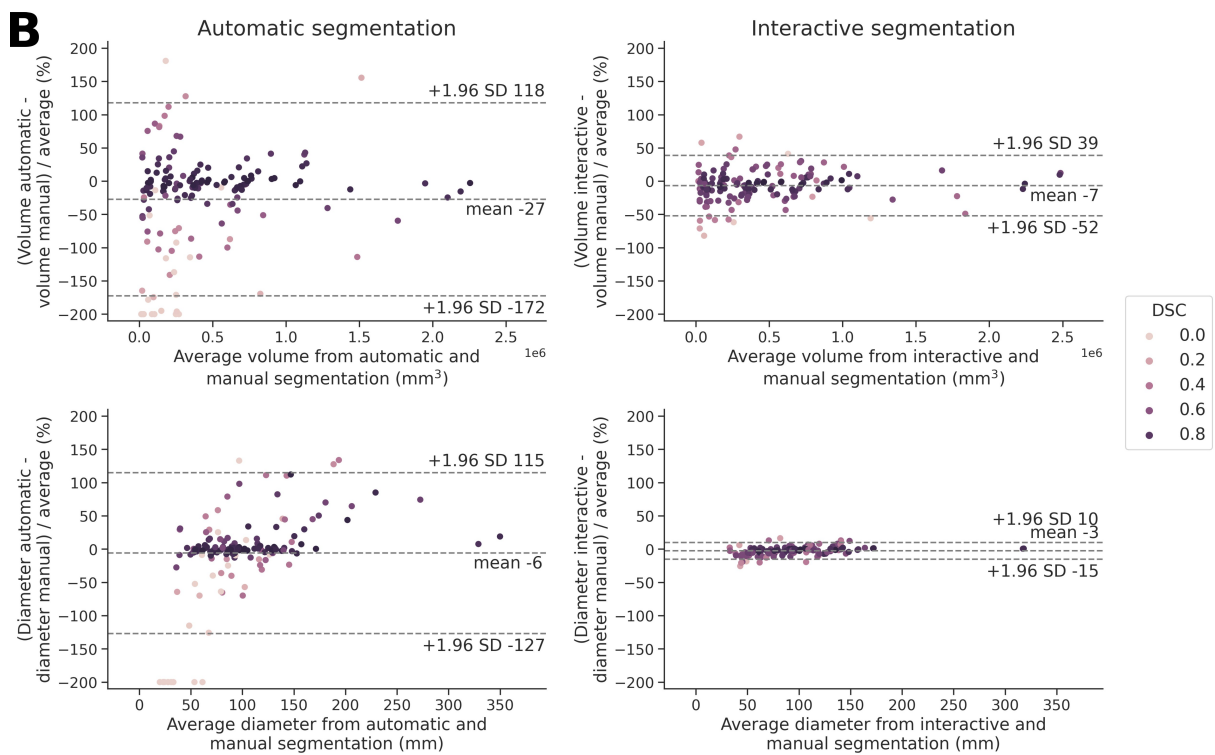

**Supplementary Figure S3**: Bland-Altman plots for percentage volume and diameter measurements from automatic and interactive compared to reference segmentations of STTs on **A)** the WORC test dataset, **B)** the TCIA test dataset. Every dot represents a patient. Patients are color-coded based on their achieved Dice Similarity Coefficient (DSC) using the respective method, i.e. automatic or interactive segmentation.

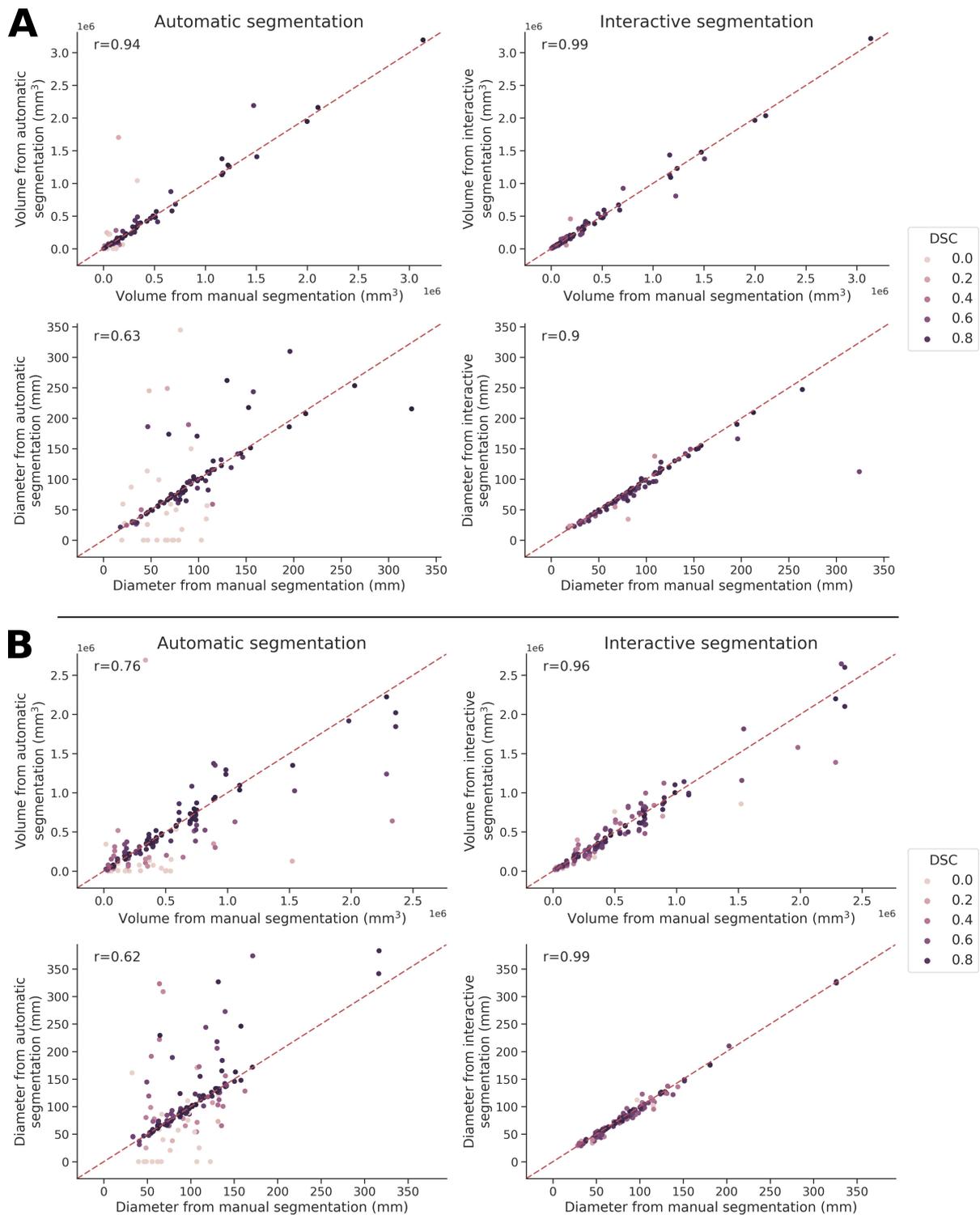

**Supplementary Figure S4:** Scatterplots for volume and diameter measurements from automatic and interactive compared to reference segmentations of STTs on **A)** the WORC test dataset, and **B)** the TCIA test dataset. Every dot represents a patient. Patients are color-coded based on their achieved Dice Similarity Coefficient (DCS) using the respective method, i.e. automatic or interactive segmentation. The dotted line represents a perfect agreement.

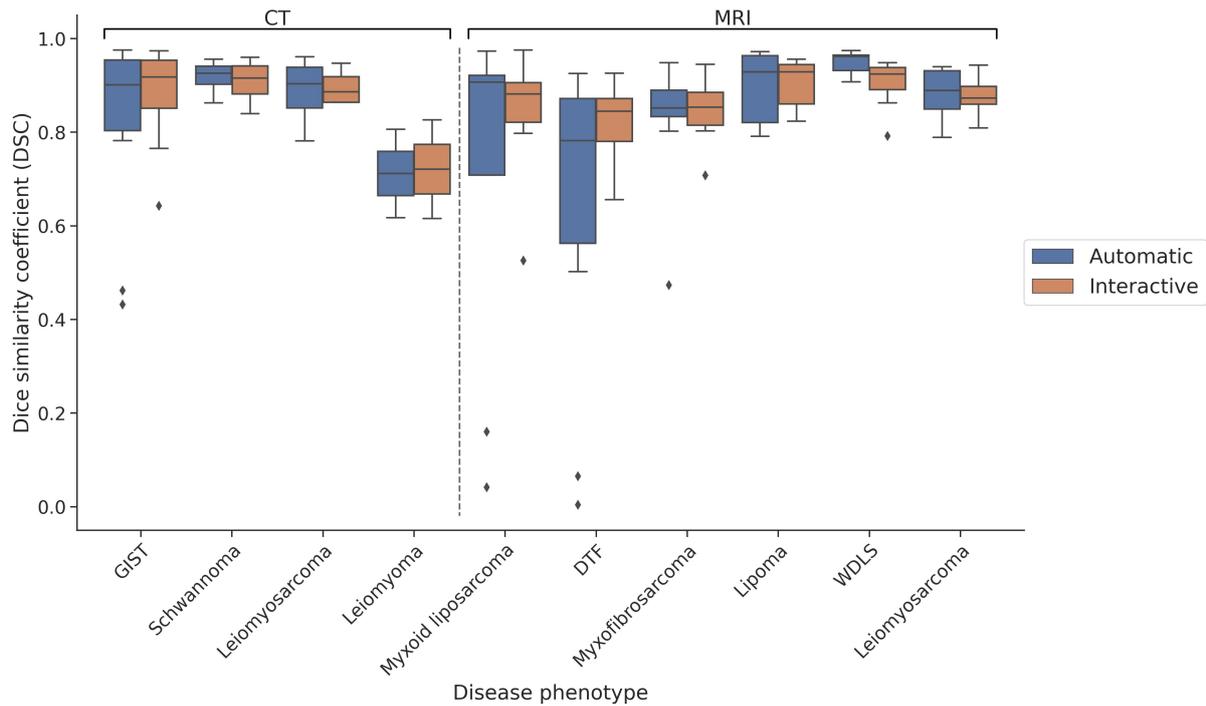

**Supplementary Figure S5:** Quantitative results from automatic and interactive segmentation of STTs on the WORC test dataset after removing the 23 outliers on which the automatic method failed to segment the correct lesion. Box-and-whisker plot, visualizing the median, quartiles, and potential outliers, for the Dice Similarity Coefficient (DSC) results of automatic (blue) and interactive (orange) segmentation methods for different phenotypes on CT and MRI. DTF: desmoid-type fibromatosis; WDLS: well-differentiated liposarcoma; and GIST: gastrointestinal stromal tumor.

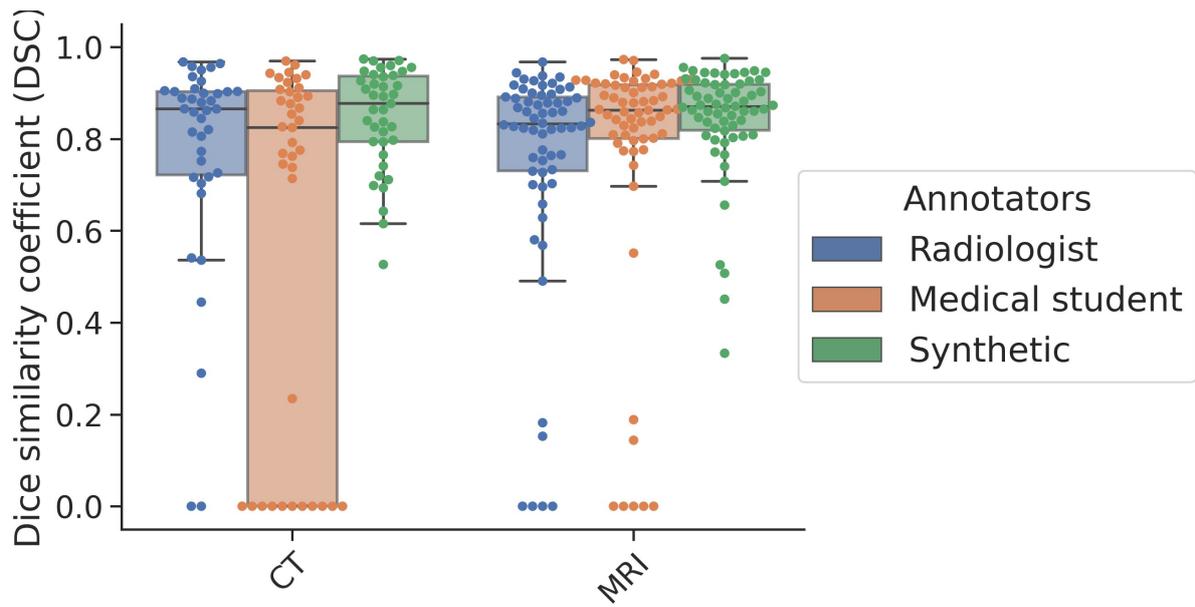

**Supplementary Figure S6**: Quantitative comparison of predicted segmentations compared to reference segmentations using the interactive segmentation pipeline for different annotators. The annotator is either the musculoskeletal radiologist (blue), medical student (orange), or synthetic annotation (green). Every dot represents a sample in the WORC test dataset.